\documentclass[twocolumn,preprintnumbers,amsmath,amssymb]{revtex4}
\usepackage{amssymb}
\usepackage[dvips]{graphicx}
\begin{document}






{\bf Reply to the comment by C. Capan and K. Behnia}

 A large Nernst signal observed  in the
normal (i.e. resistive) state of cuprates \cite{xu,cap} has been
claimed to be a signature of an unusual "normal" state  with mobile
vortexes in disagreement with the  BCS theory  and its
strong-coupling bipolaronic extension \cite{alebook}.   Such
interpretation seriously undermines many works on superconducting
cuprates, which consider the state above $T_c$ as perfectly normal
with no off-diagonal order, either long or short. We have argued
that  the vortex scenario  is impossible to reconcile with the sharp
resistive transitions at $T_c$ in high-quality cuprates, and
proposed  a theory of the Nernst signal as  a true normal state
phenomenon \cite{alezav}. The Comment \cite{cap2} claims that the
theory \cite{alezav} is incompatible with a low value of thermopower
$S$ times the Hall angle, $S\tan\Theta$ at low temperatures in
underdoped cuprates.

I believe that the criticism \cite{cap2} is faulty and misleading.
In fact, the theory \cite{alezav} did not make any general
assumption on the relative magnitudes of the Nernst sugnal $e_{y}$
and $S\tan\Theta$. We  demonstrated very good agreement between the
theory and the experimental data in overdoped La$_{1.8}$
Sr$_{0.2}$CuO$_4$, Fig.2 \cite{alezav}, where $S\tan\Theta
> e_{y}$. Here I show that the same theory
describes  well the Nernst signal and $S\tan\Theta$ also in
underdoped cuprates \cite{cap,cap2}, where $S\tan\Theta \ll e_{y}$
at low temperatures. The authors of the comment \cite{cap2} have
missed an important point of the theory \cite{alezav} i.e. that
localisation of  carriers
 below the mobility edge breaks the electron-hole symmetry
  at variance with  ordinary metals where the familiar
"Sondheimer" cancelation  makes  $e_{y}$ much smaller than
$S\tan\Theta$ because of this symmetry. The localised carrier
contribution \emph{ adds} to the contribution of itinerant carriers
to produce a large $e_{y}$, while it \emph{reduces } $S$ and
$\Theta$. Such behaviour originates in the familiar "sign" (or
"$p-n$") anomaly of the Hall conductivity of localised carriers. The
sign of their Hall effect is often $opposite$ to that of the
thermopower as observed in many amorphous semiconductors \cite{ell}
and described theoretically \cite{fri}.
\begin{figure}
\begin{center}
\includegraphics[angle=270,width=0.45\textwidth]{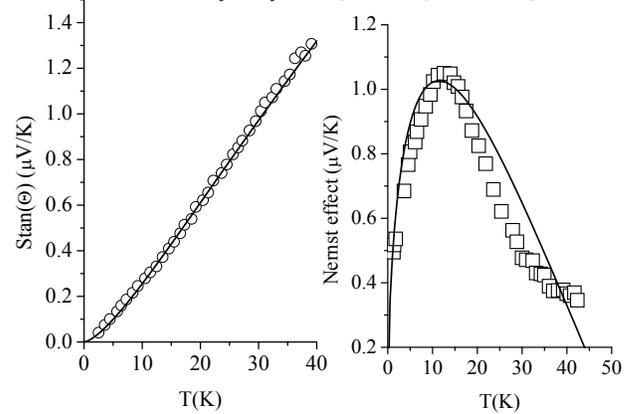}
\vskip -0.5mm \caption{$S\tan\Theta$ (circles \cite{cap2} )  and the
Nernst effect $e_y$ (squares \cite{cap})  of underdoped La$_{1.94}$
Sr$_{0.06}$CuO$_4$ at $B=12$T compared with the theory, (solid
lines).}
\end{center}
\end{figure}
Hence, at small $\Theta \ll 1$, one can write $S\tan\Theta \approx
\rho (\alpha ^{ext}_{xx}-|\alpha ^{l}_{xx}|)
(\Theta^{ext}-|\Theta^{l}|)$ and $e_{y}\approx \rho (\alpha^{ext}
_{yx}+|\alpha^{l} _{yx}|)-S\tan \Theta$, where kinetic coefficients
are expressed as the sums of extended ($ext$) and localised ($l$)
carrier contributions, $\Theta^{ext}\equiv
\sigma^{ext}_{yx}/\sigma_{xx}$, $\Theta^{l}\equiv
\sigma^{l}_{yx}/\sigma_{xx}$, and $\rho=1/\sigma_{xx}$ is the
resistivity. Clearly  these expressions can  account for a low value
of $S\tan\Theta$ compared with a large value of $e_y$  in underdoped
cuprates, where the contributuon of localised carriers is comparable
with the extended carrier contribution. To illustrate quantitative
agreement with the experiment \cite{cap,cap2} we use a textbook
result $S \sim T$ valid near the mobility edge in amorphous
semiconductors at low temperatures \cite{mott3} and the Boltzmann
scaling, $\alpha_{yx} \propto B/\rho^2$,  $\Theta \propto B/\rho$ in
the magnetic field $B$. This gives $S\tan\Theta = AT / (T_1\rho)$
and $e_{y}= A (1-T/T_1)/\rho$, where  $A$ and $T_1$ are temperature
independent. Even with these simplifications the theory describes
 both $S\tan\Theta$ and $e_y$ measured in La$_{1.94}$
Sr$_{0.06}$CuO$_4$
 \cite{cap,cap2} using a $single$ fitting parameter, $T_1=50$K,  the
experimental $\rho(T)$ \cite{cap}, and the scaling constant $A=0.78$
$\mu$V m$\Omega\cdot$ cm/K, Fig.1. The same theory \cite{alelog}
also fits nicely the insulating-like low-temperature dependence of
$\rho(T)$ revealed in high magnetic fields \cite{cap,cap2}.

To sum up, the Comment \cite{cap2} has deplorably neglected
essential parts of Ref. \cite{alezav}, which actually addressed the
low value of $S\tan\Theta$ in underdoped cuprates, the absence of a
positive $e_y$ in non-superconducting cuprates, and its nonlinear
magnetic field dependence (pages 4,3 and Fig.2 in \cite{alezav},
respectively). The coexistence of the large Nernst signal and the
insulating-like resistivity in underdoped cuprates \cite{cap,cap2}
sharply disagrees with the vortex scenario \cite{xu}, but agrees
remarkably well  with our theory \cite{alezav}.

\noindent {\bf A.S. Alexandrov}, Department of Physics, Loughborough
University, Loughborough LE11 3TU, United Kingdom.


\begin{thebibliography}{90}
\bibitem{xu}  Z.A. Xu et al., Nature (London) \textbf{406}, 486
(2000).
\bibitem{cap} C. Capan et al., Phys. Rev. Lett. {\bf
88}, 056601 (2002).
\bibitem{alebook} A.S. Alexandrov, \emph{Theory of Superconductivity: From Weak to Strong Coupling}
(IoP Publishing, Bristol, (2003)).
\bibitem{alezav} A.S. Alexandrov and V.N. Zavaritsky, Phys. Rev. Lett. {\bf 93},
217002 (2004).
\bibitem{cap2} C. Capan and K. Behnia, preceding
comment.
\bibitem{ell} S.R. Elliot, \emph{Physics of amorphous
materials}, pp. 222-225 (Longman, New York, 1983).
\bibitem{fri} L. Friedman, J. Non-Cryst.Sol. {\bf 6}, 329 (1971).
\bibitem{mott3} M. Cutler and N.F. Mott, Phys. Rev. {\bf 181}, 1336
(1969).
\bibitem{alelog} A.S. Alexandrov, Phys. Lett. A{\bf 236}, 132
(1997); cond-mat/0507268.



\end{thebibliography}
\end{document}